\documentclass[11pt, twoside]{article}
\usepackage{latexsym,atmp}

\copyrightnotice{2002}{6}{1163}{1182}
\newtheorem{theorem}{Theorem}
\newtheorem{lemma}{Lemma}[section]
\newtheorem{prop}{Proposition}[section]
\newtheorem{corol}{Corollary}[section]
\newtheorem{definition}{Definition}
\newtheorem{remark}{Remark}

\def\R{\mathbb R}
\def\S{\mathcal{S}^2}
\def\G{\mathcal{G}}

\def\S{\Sigma}
\def\pf{\noindent \emph{Proof}: }
\def\ep{\epsilon}
\def\del{\delta}
\def\sig{\sigma}
\def\gam{\gamma}
\def\lbar{\overline}
\def\norm{\|}
\def\stop{\hfill$\Box$}
\def\dt{\partial_t}
\def\dxi{\frac{\partial }{\partial x^i}}
\def\dxj{\frac{\partial }{\partial x^j}}
\def\odt{\frac{d}{dt}}
\def\pdt{\frac{\partial}{\partial t}}
\def\ods{\frac{d}{ds}}
\def\hatg{\hat{g}_\del}
\def\ms{\mathcal{M}^2(\S)}
\def\mz{\mathcal{M}^0(\S)}
\def\mf{\mathcal{M}^1(\S)}

\def\Rdel{{R_\del}}
\def\Rdeln{\Rdel_{-}}
\def\tilM{\tilde{M}}
\def\tilg{\tilde{g}_\del}
\def\mudel{\mu_\delta}
\def\tilR{\tilde{R}_\del}
\def\gi{g_{-}}
\def\go{g_{+}}
\def\sigdel{\sigma_\del}

\begin{document}

\title{Positive Mass Theorem on Manifolds admitting Corners along a 
Hypersurface}

\url{math-ph/0212025}
\vspace{-.2in}

\author{Pengzi Miao}

\vspace{-.2in}

\address{Department of Mathematics \\
Stanford University \\
Stanford, CA 94305 \\
USA}
\addressemail{mpengzi@math.stanford.edu}  

\markboth{\it Positive Mass Theorem\ldots}{\it Pengzi Miao}

\begin{abstract}
We study a class of non-smooth asymptotically flat manifolds on which metrics 
fail to be $C^1$ across a hypersurface $\Sigma$. We first give an 
approximation scheme to mollify the metric, then we show that the Positive 
Mass Theorem \cite{Sch-Yau} still holds on these manifolds if a geometric 
boundary condition is satisfied by metrics separated by $\S$.
\end{abstract}  

\section{Introduction and Statement of Results}
The well-known Positive Mass Theorem in general relativity was first proved
by R. Schoen and S.T. Yau in \cite{Sch-Yau} for smooth asymptotically flat 
manifolds with non-negative scalar curvature. 
It is interesting to know on what kind of non-smooth Riemannian
manifolds their techniques and results can be generalized. 
In this paper we study this question in a special setting where the metric 
fails to be $C^1$ across a hypersurface.

\cutpage

Let $n \geq 3$ be a dimension for which the classical PMT \cite{Sch-Yau} holds.
Let $\alpha \in (0, 1)$ be a fixed number and $M$ be an oriented 
$n$-dimensional smooth differentiable manifold with no boundary. We assume 
that there exists a compact domain $\Omega \subset M$ so that 
$M \setminus \Omega$ is diffeomorphic to $\R^n$ minus a ball and 
$ \Sigma = \partial{\Omega}$ is a smooth hypersurface in M.

\begin{definition}
{\bf A metric $\mathcal{G}$ admitting corners along $\Sigma$}
is defined to be a pair of $(g_{-}, g_{+})$, where $g_{-}$ and
$g_{+}$ are $C^{2,\alpha}_{loc}$ metrics on $\Omega$ and 
$M \setminus \lbar{\Omega}$
so that they are $C^2$ up to the boundary and they induce the same metric on
$\S$.
\end{definition}

\begin{definition} Given $\G = (g_{-}, g_{+})$, we say $\G$ is 
{\bf asymptotically flat} if the manifold 
$(M \setminus \Omega, g_{+})$ is asymptotically 
flat in the usual sense {\rm (see \cite{Sch})}.
\end{definition}    

\begin{definition} The {\bf mass} of $\G = (g_{-}, g_{+})$ is defined 
to be the  mass of $g_{+}$ {\rm (see \cite{Sch})} whenever the later 
exits.     
\end{definition}  

One of our main motivation to study such a pair $\G = (g_{-}, g_{+})$
is its implicit relation with Bartnik's quasi-local mass of the 
bounded Riemannian domain $(\lbar{\Omega}, \gi)$. It is 
generally conjectured that 
there exists a $g_{+}$ on $M \setminus \Omega$ so that $\G = (g_{-}, g_{+})$
is a minimal mass extension of $(\lbar{\Omega}, \gi)$ 
in the sense of \cite{Bartnik_local}.

Under a geometric boundary condition which originated in \cite{Bartnik}, 
we prove the following Positive Mass Theorem for $\G$.

\begin{theorem}\label{PMT1} 
Let $\G = ( g_{-}, g_{+})$ be an asymptotically flat metric admitting corners
along $\Sigma$. Suppose that the scalar curvature of $\gi$, $\go$ is
non-negative in $\Omega$, $ M \setminus \lbar{\Omega}$, and 
$$ \hspace{4cm} H(\Sigma, \gi) \geq H(\Sigma, \go), \hspace{4cm} {\rm (H)} $$
where $H(\Sigma, \gi)$ and $H(\Sigma, \go)$ represent the mean 
curvature of $\Sigma$ in $(\lbar{\Omega}, \gi)$ and $(M \setminus 
\Omega, \go)$ with respect to unit normal vectors pointing to the 
unbounded region.
Then the mass of $\G$ is non-negative. Furthermore, if $H(\S, \gi) > 
H(\S, \go)$ at some point on $\S$, $\G$ has a strict positive mass.
\end{theorem}

\begin{remark} 
Under our sign convention for the mean curvature, we have that
$H(S^{n-1}, g_o) = n-1$, 
where $S^{n-1}$ is the unit sphere in $\R^n$ and $g_o$ is the Euclidean 
metric.
\end{remark}

One direct corollary of this theorem is that the boundary behavior of a metric
$g$ on $\lbar{\Omega}$ imposes subtle restriction on the scalar curvature of 
$g$ inside $\Omega$. For instance, we have that

\begin{corol}
There does not exist a metric $g$ with non-negative scalar curvature on a 
standard ball $\lbar{B} \subset \R^n $ so that $\partial B$
is isometric to $S^{n-1}$ and the mean curvature of 
$\partial B$ in $(\lbar{B}, g)$ is greater but not equal to $n-1$.
\end{corol}

Based on the work of H. Bray and F. Finster \cite{Bray-Finster}, we have
a rigidity characterization of $\G$ when its mass is zero.

\begin{theorem}\label{PMT2} 
Let $n = 3$ and $\gi, \go$ satisfy all
the assumptions in Theorem \ref{PMT1}. If $\gi$ and $\go$ are at least 
$C^{3, \alpha}_{loc}$, then the mass of $\go$ being zero implies that 
$\gi$ and $\go$ are flat away from $\S$ and they induce the same second 
fundamental form on $\S$. Hence, $(\Omega, \gi)$ and 
$(M \setminus \lbar{\Omega}, \go)$ together can 
be isometrically identified with the Euclidean space $(\R^3, g_o)$.
\end{theorem}

To illustrate the relevance of Theorem \ref{PMT2} to the quasi-local mass 
of a bounded Riemannian domain, we mention the following corollary. 

\begin{corol}
Let $(M^3, g)$ be a manifold with non-negative scalar curvature, 
possibly with 
boundary. Let $g_\sig$ be a metric on $S^2$ so that there exist two 
isometric embeddings $\phi_1: (S^2, g_\sig) \rightarrow (\R^3, g_o)$ and 
$\phi_2: (S^2, g_\sig) \rightarrow (M^3,g)$, where
$\phi_1(S^2), \phi_2(S^2)$ each bounds a compact region 
$\Omega_1, \Omega_2$ in $\R^3$, $M^3$ that has connected 
boundary. Then, if
$$  H(\phi_2(S^2), g) \geq H(\phi_1(S^2), g_o), $$ 
$\Omega_2$ is isometric to $\Omega_1$. In particular, 
$\Omega_2$ has trivial topology.
\end{corol}

\begin{remark}
If we replace $S^2$ by an arbitrary compact surface $\S_g$ with genus 
$g \geq 1$, under the same assumption, our argument still works to show 
that the region bounded by $\phi_2(\S_g)$ is {\bf flat}.
\end{remark}

\section{Explanation of condition (H)}
In this section we give a motivation for the geometric boundary condition 
({\bf H}). One will see that it can be interpreted as a statement that the 
scalar curvature of $\G$ is distributionally non-negative across $\Sigma$.

Let $g$ be a $C^2$ metric in a tubular neighborhood of $\S$ and $\nu$ be 
a unit normal vector field to $\S$. Let $K$ be the Gaussian curvature of $\S$ 
with respect to the induced metric $g |_\S$ and $R$ be the scalar 
curvature of $g$.
Taking trace of the Gauss equation, we have that
\begin{equation}\label{Gauss}
  2K = R - 2Ric(\nu, \nu) + H^2 - {| A |}^2,  
\end{equation} 
where $Ric(\nu, \nu)$ is the Ricci curvature of $g$ along $\nu$, $H$ 
and $A$ are the mean curvature and the second fundamental form of $\S$.

Assuming that $\S$ evolves with speed $\nu$, we have the following evolution 
formula of the mean curvature
\begin{equation}\label{DH}
  D_{\nu}H = -Ric(\nu, \nu) - {|A|}^2 .
\end{equation}  
It follows from (\ref{Gauss}) and (\ref{DH}) that 
\begin{equation}\label{RH}
 R = 2K - ({|A|}^2 + H^2) - 2D_{\nu}H ,
\end{equation} 
which suggests that
$D_{\nu}H$ plays a dominant role in determining the sign of $R$ if 
$K, H$ and 
$A$ are known to be bounded. In particular,
for a metric $\G = (\gi, \go)$ with $H(\S,\gi) > H(\S, \go)$, the 
scalar curvature of $\G$ across $\S$ looks like a positive Dirac-Delta
function with support in $\S$. Hence, the spirit of 
Theorem \ref{PMT1} is that PMT still holds even if the scalar curvature is 
only assumed to be distributionally non-negative across $\S$. 

\begin{remark} 
The geometric boundary condition {\em ({\bf H})} was 
first introduced by R. Bartnik in {\em \cite{Bartnik}}, 
where he suggested the static metric extension conjecture for a bounded 
domain in a time-symmetric initial data set.
\end{remark}

\section{Smoothing $\mathcal{G}$ across $\S$}
Given $\G=(g_{-}, g_{+})$ on $M$, we want to approximate $\G$ by 
metrics which are $C^2$ across $\S$. 

First, we use the Gaussian coordinates of $\S$ to modify the 
differential structure on $M$ so that $\G$ becomes a continuous metric 
across $\S$. Let $U_{-}^{2\ep}$ be a $2\ep$-tubular neighborhood of
$\S$ in $(\lbar{\Omega}, g_{-})$ for some $\ep>0$. Let 
$$ \Phi_{-}: \S \times (-2\ep, 0] \longrightarrow U_{-}^{2\ep} $$
be a diffeomorphism so that the pull back metric $\Phi_{-}^{*}(g_{-})$ has 
the form 
$
\Phi_{-}^{*}(g_{-}) = {g_{-}}_{ij}(x,t)dx^idx^j + dt^2  , 
$
where $t$ is the coordinate for $(-2\ep, 0]$, $(x^1, \ldots, x^{n-1})$ are 
local coordinates for $\S$ and $i,j$ runs through $1, \ldots, n-1$. 
Similarly, we have 
$ \Phi_{+}: \S \times [0, 2\ep) \longrightarrow U_{+}^{2\ep}, $
where $U^{2\ep}_{+}$ is a 2$\ep$-tubular neighborhood of $\S$ in 
$(M \setminus \Omega, \go)$ and  
$\Phi_{+}^{*}(g_{+}) = {g_{+}}_{ij}(x,t)dx^idx^j + dt^2. $
Identifying $U = U_{-}^{2\ep} \cup U_{+}^{2\ep}$ with $\S \times (-2\ep,
2\ep)$, we define $\tilde{M}$ to be a possibly new differentiable manifold 
with the background topological space $M$ and the differential structure 
determined by the open covering $\{\Omega, M \setminus \lbar{\Omega}, U\}$.
Since $g_{-}|_\S = g_{+}|_\S$, $\G$ becomes a continuous metric 
$g$ on $\tilde{M}$. Inside $U = \S \times (-2\ep, 2\ep)$, we have that
\begin{equation}\label{pathg}
 g = g_{ij}(x,t)dx^idx^j + dt^2, 
\end{equation}
where $ g_{ij}(x,t) = {g_{-}}_{ij}(x,t) $ when $ t\leq 0 $ and 
$ g_{ij}(x,t) = {g_{+}}_{ij}(x,t) $ when $ t\geq 0 $.
   
Second, we mollify the metric $g$ inside $U$. Let $i \in
\{0,1, 2\}$, we define $\mathcal{S}^i(\S)$ to be the Banach space of
$C^i$ symmetric $(0,2)$ tensors on $\S$ equipped with the usual $C^i$ norm
and $\mathcal{M}^i(\S)$ to be the open and convex subset of 
$\mathcal{S}^i(\S)$ consisting of $C^i$ metrics. 
By (\ref{pathg}) we have a well defined path in each $\mathcal{M}^i(\S)$,
\begin{equation}
 \gamma: (-2\ep,2\ep) \longrightarrow \ms \hookrightarrow \mf \hookrightarrow
 \mz 
\end{equation}  
where 
\begin{equation}
\gamma(t) = g_{ij}(x,t)dx^idx^j.
\end{equation} 
By assumption, $\gamma$ is a continuous path in $\ms$ and a piecewise $C^1$ 
path in $\mf$. Hence, there exists $L > 0$ depending 
only on $\G$ such that
\begin{equation}\label{Lip}
 \| \gamma(t) - \gamma(s) \|_{\mf} \leq  L |t-s|, \ \  \forall s, t \in 
[-\frac{3}{2}\ep,\frac{3}{2}\ep] .
\end{equation} 
We choose $\phi(t) \in {C_{c}}^\infty([-1, 1])$ to be a standard mollifier 
on $\R^1$ such that
\begin{equation}
  0 \leq \phi \leq 1
  \mathrm{\ \ and\ \ } \int^{1}_{-1} \phi(t) dt = 1 . 
\end{equation}    
Let $\sigma(t) \in {C_{c}}^\infty([-\frac{1}{2}, \frac{1}{2}])$ be anther 
cut-off function such that
\begin{equation}
  \left\{ \begin{array}{ll}
           0 \leq \sigma(t) \leq \frac{1}{100} &  t \in \R^1 \\
          \sigma(t) = \frac{1}{100} &  |t| < \frac{1}{4}  \\
          0< \sigma(t) \leq \frac{1}{100} & \frac{1}{4} <
          |t| < \frac{1}{2} . 
         \end{array} 
 \right.
\end{equation}
Given any $0 < \delta \ll \ep$, let
\begin{equation}
    \sigma_{\del}(t) = \del^2 \sig(\frac{t}{\del})
\end{equation}
and we define
\begin{eqnarray}\label{molligam}
    \gam_{\del}(s) & = & \int_{\R^1} \gam(s - \sig_{\del}(s)t) \phi(t)
    dt, \hspace{1.1cm} s \in (-\ep, \ep)  \nonumber \\
    & = & \left\{ \begin{array}{ll}
                   \int_{\R^1} \gamma(t) \left( \frac{1}{\sigdel(s)}
                   \phi(\frac{s-t}{\sigdel(s)}) \right) dt, & 
                    \sigdel(s) > 0 \\
                   \gamma(s), & \sigdel(s) = 0 ,
                 \end{array}
          \right. 
\end{eqnarray}
where the integral takes place in $\mathcal{S}^0(\S)$. By the convexity of 
$\mathcal{M}^0(\S)$ in $\mathcal{S}^0(\S)$, $\gamma_\del$ is a path in 
$\mathcal{M}^0(\S)$. We have the following elementary lemmas concerning the 
property of $\gamma_\del$ and its relation with $\gamma$. 

\begin{lemma} \label{c12gamma}
 $\gam_{\del}(s)$ is a  $C^2$ path in $\mz$ and a $C^1$ path in $\mf$.
\end{lemma}  
\pf Let $i \in \{1, 2\}$. The fact that 
$\gamma:(-\ep, \ep) \longrightarrow \mathcal{M}^{2-i}(\S)$ is $C^i$ away from 
0 implies that
$\gam_\del: (-\ep, \ep) \longrightarrow \mathcal{M}^{2-i}(\S)$ is $C^i$ away 
from $[-\frac{\del^2}{100},\frac{\del^2}{100}]$. For 
$s \in (-\frac{\del}{4}, \frac{\del}{4})$, 
$\sig_{\del}(s) = \frac{\del^2}{100}$ and (\ref{molligam}) turns into  
\begin{equation}\label{cmollify}
     \gam_\del(s)  =  \int_{\R^1} \gam(t) \left( 
     \frac{1}{(\frac{\del^2}{100})}
    \phi(\frac{s-t}{(\frac{\del^2}{100})}) \right) dt ,
\end{equation}    
which is the standard mollification of $\gam$ by $\phi$ with a
constant scaling factor $\frac{\del^2}{100}$. Hence, 
$\gam_\del(s):(-\ep, \ep) \longrightarrow \mathcal{M}^{2-i}(\S)$ is smooth in 
$ (-\frac{\del}{4}, \frac{\del}{4})$. \stop

\begin{lemma}\label{c0gamma}
$\gam_{\del}(s)$ is a $C^0$ path in $\ms$ which is uniformly close to 
$\gamma$ and agrees with $\gamma$ outside $(-\frac{\del}{2},\frac{\del}{2})$.
\end{lemma}
\pf The continuity of $\gamma_\del: (-\ep, \ep) \longrightarrow \ms$ follows
directly from that of $\gamma$. The estimate
\begin{eqnarray}\label{c2close} 
 \norm \gam_\del(s) - \gam(s) \norm_{\ms} & = & \left\|
 \int_{\R^1} \left(  \gam(s - \sig_{\del}(s)t) - \gam(s) \right)
 \phi(t) dt \right\|_{\ms} \nonumber\\ 
 & \leq &  \int_{\R^1} \| \gam(s - \sig_{\del}(s)t) - \gam(s) \|_{\ms}
 \phi(t) dt 
\end{eqnarray} 
shows it is uniformly close to $\gamma$. Finally, 
$\sig_{\del}(s) = 0$ for $|s| > \frac{\del}{2}$ implies 
\begin{equation} \gam_{\del}(s) = \int_{\R^1} \gam(s) \phi(t) dt  =
  \gam(s).
\end{equation}  \stop   

\begin{lemma}\label{delestimate}
$\| \gamma_\del(s) - \gamma(s) \|_{\mz} \leq L \del^2, \mathrm{\ for\ }
s \in (-\ep, \ep)$.
\end{lemma}
\pf It follows from (\ref{Lip}) that 
\begin{eqnarray} 
  \norm \gam_\del(s) - \gam(s) \norm_{\mz} 
  & \leq &  \int_{\R^1} \| \gam(s - \sig_{\del}(s)t) - \gam(s) \|_{\mz}
  \phi(t) dt \nonumber \\
  & \leq & L \del^2. 
\end{eqnarray}  \stop

Now we define
\begin{equation} \label{gdel}
 g_\del = \left\{
            \begin{array}{ll}
              {\gam_\del}(t) + dt^2, & (x,t) \in \S \times
              (-\ep, \ep) \\ 
              g, & (x,t) \notin \S \times (-\ep, \ep) .
            \end{array}
           \right. 
\end{equation}
Lemma \ref{c12gamma}, \ref{c0gamma} and \ref{delestimate} imply 
that $g_\del$ is a globally $C^2$ metric on $\tilde{M}$ which agrees with $g$ 
outside a strip region $\S \times (-\frac{\del}{2}, \frac{\del}{2})$ and
is uniformly close to $g$ in $C^0$ topology.

Next, we proceed to estimate the scalar curvature of $g_\del$. We 
use the notations defined in section $2$ with a lower index $\del$ to denote 
the corresponding quantities of $g_\del$.
By (\ref{gdel}), the vector field $\frac{\partial}{\partial t}$ is
perpendicular to the slice $\S \times \{ t \}$ for each $t \in (-\ep, \ep)$. 
Hence, inside $\S \times (-\ep, \ep)$, (\ref{RH}) shows that  
\begin{equation}\label{RHdel}
  R_\del(x,t) = 2K_\del(x,t) - ({|A_\del(x,t)|}^2 + H_\del(x,t)^2) - 2
  \frac{\partial}{\partial t}H_\del(x,t).
\end{equation}
We will estimate each term on the right of (\ref{RHdel}). First we note 
that $K_\del(x,t)$ is determined only by $\gamma_\del(t)$, Lemma 
\ref{c0gamma} then implies that $K_\del(x,t)$ is bounded by constants 
depending only on $\gamma: (-\ep,\ep) \longrightarrow \ms$. 

To estimate $A_\del(x,t)$ and $H_\del(x,t)$, we compute the first order
derivative of $\gamma_\del: (-\ep, \ep) \longrightarrow \mz$ because of the 
definition
\begin{equation}
   {A_\del}_{ij}(x,t)  =  <\dxj, \nabla^\del_{\dxi} \dt> =
     \frac{1}{2} \frac{\partial}{\partial t} {\gam_\del}_{ij}(x,t).
\end{equation} 
By (\ref{molligam}) we have that
\begin{equation}\label{dA}
  \pdt {\gamma_\del}_{ij}(x,t) =
  \pdt \int_{\R^1} \gam_{ij}(t - \sig_{\del}(t)s)
  \phi(s) ds.
\end{equation}  
When $|t| > \frac{\del^2}{100}$, (\ref{dA}) gives that
\begin{eqnarray}
  \pdt {\gam_\del}_{ij}(x,t) & =  & \int_{\R^1}
  \odt \{ \gam_{ij}(t - \sig_{\del}(t)s) \} \phi(s) ds  \nonumber \\
  & = & \int_{\R^1} \gam^\prime_{ij}(t - \sig_\del(t)s) \{ 1 -
  s\del\sig^\prime(\frac{t}{\del}) \} \phi(s) ds .
\end{eqnarray}
When $|t| < \frac{\del}{4}$, (\ref{cmollify}) implies that
\begin{eqnarray}
 \pdt {\gam_\del}_{ij}(x,t) & = & \pdt \int_{\R^1} \gam_{ij}(s) \left\{
  \frac{100}{\del^2} \phi ( \frac{100(t-s)}{\del^2} ) \right\} ds 
  \nonumber \nonumber \\
  & = & \int_{\R^1} \gam_{ij}(s) \odt \left\{\frac{100}{\del^2}
  \phi(\frac{100(t-s)}{\del^2}) \right\} ds \nonumber \\
  & = & (-1)\int_{\R^1} \gam_{ij}(s) \ods \left\{\frac{100}{\del^2}
  \phi(\frac{100(t-s)}{\del^2}) \right\} ds \nonumber \\ 
  & = & (-1) \int^0_{-\infty}\gam_{ij}(s) \ods \left\{\frac{100}{\del^2}
  \phi(\frac{100(t-s)}{\del^2}) \right\} ds \nonumber \\
  & + & (-1) \int^\infty_0 \gam_{ij}(s) \ods \left\{\frac{100}{\del^2}
  \phi(\frac{100(t-s)}{\del^2}) \right\} ds .
\end{eqnarray}
Integrating by parts and considering the fact  $\gam(t)$ is continuous at $0$,
 we have that
\begin{eqnarray}
  \pdt {\gam_\del}_{ij}(x,t) & = & \int_{\R^1}
   \gam_{ij}^\prime(s) \left\{\frac{100}{\del^2}
  \phi(\frac{100(t-s)}{\del^2}) \right\} ds \nonumber \\
  & = & \int_{\R^1} \gam^\prime_{ij}(t - \sig_\del(t)s) \phi(s)ds .
\end{eqnarray}
Therefore, for every $t \in (-\ep, \ep)$, we have that  
\begin{equation}\label{1dg}
  \frac{\partial}{\partial t} {\gam_\del}_{ij}(x,t) = \int_{\R^1}
   \gam^\prime_{ij}(t - \sig_\del(t)s) \left\{ 1 -
   s{\del\sig^\prime(\frac{t}{\del})} \right\} \phi(s) ds,
\end{equation}
which shows that $A_\del(x,t)$ is bounded by 
constants depending only on $\gamma: (-\ep, \ep) \longrightarrow 
\mathcal{M}^0(\S)$. Since $H_\del(x,t) = g^{ij}_\del {A_\del}_{ij} $,  
Lemma \ref{c0gamma} and (\ref{1dg}) also imply that $H_\del(x,t)$ is bounded 
by constants depending only on $\gamma: (-\ep, \ep) \longrightarrow \mz$.

To estimate $\pdt H_\del(x,t)$, we need to compute the second order
derivative of ${\gam_\del}: (-\ep,\ep) \longrightarrow \mz$. A similar 
calculation as above gives that, for $|t| > \frac{\del^2}{100}$, 
\begin{eqnarray}\label{outdh}
    \frac{\partial^2}{\partial^2 t} {\gam_\del}_{ij}(x,t) & = & \int_{\R^1}
   \gam^{\prime\prime}_{ij}(t - \sig_\del(t)s) \left\{ 1 -
   s{\del\sig^\prime(\frac{t}{\del})} \right\}^2 \phi(s) ds + \nonumber \\
   &  & \int_{\R^1}
   \gam^\prime_{ij}(t - \sig_\del(t)s) \left\{  -
   s{\sig^{\prime\prime}(\frac{t}{\del})} \right\} \phi(s) ds
\end{eqnarray} 
and, for $|t| < \frac{\del}{4}$, 
\begin{eqnarray}\label{indh}
    \frac{\partial^2}{\partial^2 t} {\gam_\del}_{ij}(x,t) & = &
    \int_{\R^1} \gam^{\prime\prime}_{ij}(t - \sig_\del(t)s) \phi(s)
    ds + \nonumber \\
    & & \left\{ {g_+}_{ij}^\prime(0) -{{g_-}_{ij}}^\prime(0)
    \right\} \left\{ \frac{100}{\del^2}\phi(\frac{100t}{\del^2}) \right\}. 
\end{eqnarray} \\
Since
\begin{equation} \label{defDH}
  \pdt H_\del(x, t) = \pdt \left\{ {g_\del}^{ij}(x, t) \right\} 
                         {A_\del}_{ij}(x, t) + {g_\del}^{ij}(x, t) 
   \pdt {A_\del}_{ij}(x, t) ,
\end{equation}
(\ref{outdh}), (\ref{1dg}) and Lemma \ref{c0gamma} imply that, outside 
$\S \times [-\frac{100}{\del^2},\frac{100}{\del^2}]$,  $\pdt H_\del(x,t)$ is
bounded by constants only depending on 
$\gamma:(-\ep,\ep) \longrightarrow \mz$.
On the other hand, inside $\S \times [-\frac{\del^2}{100}, 
\frac{\del^2}{100}]$, (\ref{indh}) and (\ref{defDH}) show that
\begin{eqnarray}\label{signDH}
  \pdt {H_\del}(x, t) & = & \pdt \left\{ {g_\del}^{ij}(x, t) \right\}
  {A_\del}_{ij}(x, t) + \nonumber \\
  & & \frac{1}{2} {g_\del}^{ij}(x, t)  \left\{ \int_{\R^1}
  \gam^{\prime\prime}_{ij}(t - \sig_\del(t)s) \phi(s) ds   \right\} +
  \nonumber\\
  & & \frac{1}{2} {g_\del}^{ij}(x, t) 
   \left\{ {g_+}_{ij}^\prime(0) -
  {g_-}_{ij}^\prime(0) \right\} \left\{
  \frac{100}{\del^2}\phi(\frac{100t}{\del^2}) \right\}.  
\end{eqnarray}
The first two terms on the right are bounded by constants depending only on 
$\gamma:(-\ep,\ep) \longrightarrow \mz$. For the third one, we rewrite it as 
\begin{eqnarray}\label{delandh}
 \frac{1}{2} \left\{ {g_\del}^{ij}(x, t) - g^{ij}(x,0) \right\} 
 \left\{ {g_+}_{ij}^\prime(0) - {g_-}_{ij}^\prime(0) \right\} 
 \left\{\frac{100}{\del^2}\phi(\frac{100t}{\del^2}) \right\} \nonumber \\
 +  \left\{ H(\S, g_{+})(x) - H(\S, g_{-})(x) \right\} \left\{
 \frac{100}{\del^2}\phi(\frac{100t}{\del^2}) \right\} .
\end{eqnarray}  
By  (\ref{Lip}), Lemma \ref{delestimate} and the fact 
$|t| \leq \frac{\del^2}{100}$ , we have that
\begin{eqnarray}
  | {g_\del}^{ij}(x, t) - g^{ij}(x,0)| & \leq & | {g_\del}^{ij}(x, t) -
  g^{ij}(x,t)| + |{g}^{ij}(x, t) - g^{ij}(x,0)| \nonumber \\
  & \leq & C L\del^2 + C L\del^2,    
\end{eqnarray} 
where $C>0$ only depends on $\G$.
Hence, we conclude that
\begin{equation}\label{boundDH}
\pdt {H_\del}(x, t)  =  O(1) + \left\{ H(\S, g_{+})(x) - H(\S, g_{-})(x) 
\right\} \left\{ \frac{100}{\del^2}\phi(\frac{100t}{\del^2}) \right\}
\end{equation} 
inside $\S \times [-\frac{\del^2}{100},\frac{\del^2}{100}]$, 
where $O(1)$ represents bounded quantities with bounds depending only 
on $\G$. 

We summarize the features of $\{ g_\del \}$ in the following proposition.

\begin{prop}\label{smoothcorner}
Let $\G = (g_{-}, g_{+})$ be a metric admitting corners along $\S$. Then 
$\exists$ a family of $C^2$ metrics $\{ g_\del \}_{0 < \del \leq \del_0}$ 
$on$ $\tilde{M}$ so that $g_\del$ is uniformly close to $g$ on $\tilde{M}$, 
$g_\del = g$ outside $\S \times (-\frac{\del}{2}, \frac{\del}{2})$ and 
the scalar curvature of $g_\del$ satisfies 
\begin{eqnarray}
  R_\del(x,t) &  = & O(1), \ \ for\  (x,t) \in \S \times \{\frac{\del^2}{100}
  < |t| \leq \frac{\del}{2}\} \\
  R_\del(x,t) &  = & O(1) + \{H(\S, \gi)(x) - H(\S, \go)(x) \} \left\{ 
  \frac{100}{\del^2}\phi(\frac{100t}{\del^2}) \right\}, \nonumber \\
  & & for \  (x,t) \in \S \times [-\frac{\del^2}{100},\frac{\del^2}{100}] ,
\end{eqnarray}
where $O(1)$ represents quantities that are bounded by constants depending 
only on $\G$, but not on $\del$.
\end{prop}

In case $H(\S, g_{-}) \equiv H(\S, g_{+})$, the following corollary 
generalizes a reflecting argument used by H. Bray in his proof of the 
Riemannian Penrose Inequality 
\cite{Bray-Pen}.

\begin{corol}
Given $\G = (g_{-}, g_{+})$, if 
$H(\S, g_{-}) \equiv H(\S, g_{+})$,
then $\exists$ a family of $C^2$ metrics 
$\{ g_\del \}_ {0 < \del \leq \del_0}$
on $\tilde{M}$ so that $g_\del$ is uniformly close to $g$ on $\tilde{M}$, 
$g_\del = g$ outside $\S \times (-\frac{\del}{2}, \frac{\del}{2})$ and the 
scalar curvature of $g_\del$ is uniformly bounded inside 
$ \S \times [-\frac{\del}{2}, \frac{\del}{2}]$ with bounds depending  
only on $\G$, but not on $\del$.  
\end{corol}

\section{Proof of Theorem \ref{PMT1}}
We fix the following notations.
Given a function $f$, we let $f_{+}$ and $f_{-}$ denote the positive and
negative part of $f$, i.e. $f = f_{+} - f_{-} $ and $|f| =  f_{+} + f_{-}$.
Given a metric $g$, we define the conformal Laplacian of $g$ to be 
$ L_g(u) = \triangle_g u - c_n R(g) u$, where $c_n = \frac{n-2}{4(n-1)}$ 
and $R(g)$ is the scalar curvature of $g$. The mass of $g$ will be 
denoted by $m(g)$ if it exists. Finally, we let $C_0, C_1, C_2, \ldots$ 
represent constants depending only on $\G$.

Throughout this section, we assume that $R(\gi), R(\go) \geq 0$ in $\Omega$, 
$M \setminus \lbar{\Omega}$, and 
$ H(\S, g_{-})(x) \geq H(\S, g_{+})(x)$ for all $x \in \S$.

\subsection{Conformal Deformation}
We want to modify $\{g_\del\}$ on $\tilde{M}$ to get $C^2$ metrics with
non-negative scalar curvature. For that purpose we use conformal deformation.
The following fundamental lemma is due to Schoen and Yau. Interested readers
may refer to \cite{Sch-Yau} for a detailed proof.

\begin{lemma}\label{conf} {\rm \cite{Sch-Yau}}
Let g be a $C^2$ asymptotically flat metric on $\tilde{M}$ and $f$ be a 
function that has the same decay rate at $\infty$ as $R(g)$, then $\exists$ a
number $\ep_0 > 0$ depending only on the $C^0$ norm of $g$ and the  
decay rate of $g$, $\partial g$ and $\partial \partial g$ at $\infty$ 
so that if 
\begin{equation} \label{smallep}
\left\{ \int _{\tilM} {| f_{-} |}^{\frac{n}{2}} \ dg \right\} ^{\frac {2}{n}} 
< \ep_0 ,
\end{equation}
then 
\begin{equation}\label{confequa} 
 \left\{ \begin{array}{rrr}
            {\triangle}_{g} u - c_n f u & = & 0 \\
            \lim_{x \rightarrow \infty} u & = & 1
           \end{array}
 \right. 
\end{equation}
has a $C^2$ positive solution $u$ defined on $\tilM$ so that
$u = 1 + \frac{A}{|x|^{n-2}} + \omega $
for some constant A and some function $\omega$, where 
$\omega = O(|x|^{1-n})$ and $\partial \omega = O(|x|^{-n})$. 
\end{lemma} 

For each $\del$, we consider the following equation  
\begin{equation}\label{confu} 
 \left\{ \begin{array}{rrl}
            {\triangle}_{g_\del} u_\del + c_n {R_\del}_{-} u_\del & = & 0 \\
            \lim_{x \rightarrow \infty} u_\del & = & 1 \  .
           \end{array} 
 \right. 
\end{equation}
It follows from Proposition \ref{smoothcorner} and assumptions on 
$R(\gi)$ and $R(\go)$ that 
\begin{equation} \label{boundRneg}
 \left\{ \begin{array}{ll}
          {R_\del}_{-} =  0, & \mathrm{\ outside\ } 
          \S \times [-\frac{\del}{2},
          \frac{\del}{2}]   \\
          | {R_\del}_{-} | \leq C_0, & \mathrm{\ inside\ } \S \times 
          [-\frac{\del}{2}, \frac{\del}{2}] \ . 
         \end{array}
 \right.
\end{equation} 
Therefore, (\ref{smallep}) holds with $f$ and
$g$ replaced by ${-R_\del}_{-}$ and $g_\del$, for sufficiently small $\del$. 
We note that $\ep_0$ can be chosen to be independent on $\del$ because of
Proposition \ref{smoothcorner}. Hence the solution to (\ref{confu}) 
exists by Lemma \ref{conf}. We have the following $L^{\infty}$ and 
$C^{2, \alpha}$ estimate for $\{u_\del\}$. 

\begin{prop}\label{supestimate}
$\lim_{\del \rightarrow 0} \| u_\del - 1 \|_{L^\infty(\tilM)} = 0$ and 
$\norm u_\del \norm_{C^{2, \alpha}(K)} \leq C_K$. Here $K$ is any compact set 
in ${\tilM \setminus \S}$ and $C_K$ only depends on $g$ and $K$.
\end{prop}

\pf It suffices to obtain the $L^\infty$ estimate of $| u_\del - 1 |$ 
because, once it is established, the $C^{2, \alpha}$ estimate will follow 
directly from the fact $ \triangle_{g} u_\del = 0 $ outside 
$\S \times [-\frac{\del}{2}, \frac{\del}{2}]$ and the standard Schauder 
theory. Let $w_\del = u_\del - 1$, we have that
\begin{equation}\label{eofw}
{\triangle}_{g_\del} w_\del +  c_n \Rdel_{-} w_\del  = -c_n \Rdel_{-} 
\end{equation}
where $w_\del = \frac{A_\del}{|x|^{n-2}} + \omega_\del$ for some constant 
$A_\del$ and some function $\omega_\del$ with the decay 
rate in Lemma \ref{conf}.
Multiply (\ref{eofw}) by $w_\del$ and integrate over $\tilM$,
\begin{equation} \int_{\tilM} (w_\del {\triangle}_{g_\del} w_\del + c_n 
\Rdel_{-} {w_\del}^2) \ d{g_\del} = \int_{\tilM} - c_n \Rdel_{-} {w_\del} \ 
d{g_\del}.
\end{equation}
Integrating by parts and using H$\ddot{o}$lder Inequality, we have that
\begin{eqnarray}\label{IandH}
 \int_{\tilM} {| \nabla_{g_\del} w_\del |}^2 d{g_\del}  & \leq & 
 c_n {\left( \int_{\tilM} | \Rdeln |^\frac{n}{2} \ d{g_\del} \right)}^
 {\frac{2}{n}} {\left( \int_{\tilM} {w_\del}^\frac{2n}{n-2} \ d{g_\del} 
 \right)}^{\frac {n-2}{n}} \nonumber \\
 & & + c_n {\left( \int_{\tilM} | \Rdeln |^\frac{2n}{n+2} \ d{g_\del} 
 \right)}^
 {\frac {n+2}{2n}} {\left( \int_{\tilM} {w_\del}^\frac{2n}{n-2} \ d{g_\del} 
 \right)}^{\frac {n-2}{2n}}.
\end{eqnarray}
On the other hand, the Sobolev Inequality gives that   
\begin{equation}\label{Sob}
 {\left( \int_{\tilM} {w_\del}^\frac{2n}{n-2} \ d{g_\del} \right)}^
 {\frac{n-2}{n}} 
 \leq C_\del \int_{\tilM} {| \nabla_{g_\del} w_\del |}^2 \ d{g_\del} , 
\end{equation} 
where $C_\del$ denotes the Sobolev Constant of the metric $g_\del$. 
It follows from (\ref{IandH}), (\ref{Sob}) and the elementary inequality 
$ ab \leq \frac{a^2}{2} + \frac{b^2}{2} $ that
\begin{eqnarray}\label{abw}
{\left( \int_{\tilM} {w_\del}^\frac{2n}{n-2} \ d{g_\del} \right)}^
{\frac{n-2}{n}} & \leq & C_\del c_n {\left( \int_{\tilM} {| \Rdeln |}^
\frac{n}{2} \ d{g_\del} 
\right)}^{\frac{2}{n}} {\left( \int_{\tilM} {w_\del}^\frac{2n}{n-2} \ 
d{g_\del} \right)}^\frac{n-2}{n} \nonumber \\ 
& & + \frac{1}{2}{C_\del}^2 {c_n}^2 {\left( \int_{\tilM} {| \Rdeln |}^
\frac{2n}{n+2} \ d{g_\del} \right)}^\frac{n+2}{n} \nonumber \\ 
& & + \frac{1}{2}{\left(\int_{\tilM} {w_\del}^\frac{2n}{n-2} \ d{g_\del} 
\right)}^\frac{n-2}{n}.
\end{eqnarray}
We note that Proposition \ref{smoothcorner} implies that $C_\del$ is 
uniformly close to the Sobolev Constant of $g$. Hence, for sufficiently small 
$\del$, (\ref{abw}) gives that
\begin{equation} \label{smallo}
{\left( \int_{\tilM} {w_\del}^\frac{2n}{n-2} \ d{g_\del} \right)}^
{\frac{n-2}{n}} \leq C {\left( \int_{\tilM} {| \Rdeln |}^\frac{2n}{n+2} \ 
d{g_\del} \right)}^\frac{n+2}{n} 
 = o(1), \mathrm{\ \ as\ } \del \rightarrow 0 .
\end{equation}
This $L^\frac{2n}{n-2}$ estimate and (\ref{eofw}) imply the 
supremum estimate for $w_\del$
\begin{eqnarray} 
\sup_{\tilM} |w_\del| & \leq & C \left\{ \left( \int_{\tilM} 
{w_\del}^\frac{2n}{n-2} \ d{g_\del} \right)^\frac{n-2}{2n} + \left( 
\int_{\tilM} | \Rdeln |^\frac{n}{n-2} d g_{\del} \right)^\frac{n-2}{n} 
\right\} \nonumber \\ & = & o(1) \mathrm{\  \ as\ } \del 
\rightarrow 0 
\end{eqnarray} 
by the standard linear theory(Theorem 8.17 in\cite{G-Tr}). \stop

Now we define
\begin{equation}
\tilg = {u_\del}^\frac{4}{n-2} g_\del .
\end{equation}
It follows from Proposition \ref{supestimate} that, 
passing to a subsequence, 
$\{ \tilg \}$ converges to $g$ in $C^0$ topology on $\tilde{M}$ and in 
$C^2$ topology on compact sets away from $\S$. By the conformal 
transformation formulae of scalar curvature \cite{Sch}, we also have that 
\begin{equation}
 \tilde{R}_\del = -c^{-1}_n u^{-(\frac{n+2}{n-2})}_\del L_{g_\del}(u_\del) 
  =  u^{\frac{4}{2-n}}_\del{R_\del}_{+} \geq 0 ,
\end{equation} 
where $ \tilde{R}_\del $ represents the scalar curvature of $\tilg$.

\begin{lemma}\label{massconverge}
The mass of $\tilde{g}_\del$ converges to the mass of $\G$.
\end{lemma}

\pf A straightforward calculation using the definition of mass reveals that
\begin{equation} \label{m1}
m(\tilg) = m(g_\del) + (n-1) A_\del ,
\end{equation}
where $A_\del$ is given by the expansion
$u_\del(x) = 1 + A_\del |x|^{2-n} + O(|x|^{1-n})$.
Applying integration by parts to (\ref{confu}) multiplied by $u_\del$, we have 
that
\begin{equation} \label{A1}
 (2-n)\omega_n A_\del =  \int_{\tilM} \left[ | \nabla_{g_\del} u_\del |^2 - 
 c_n {R_\del}_{-} {u_\del}^2 \right] d g_\del ,
\end{equation}
where $\omega_n$ is the volume of the $n-1$ dimensional unit sphere 
in $\R^n$. It follows from that (\ref{m1}) and (\ref{A1}) that
\begin{equation} \label{massdifference}
m(g_\del) = m(\tilg) + \frac{n-1}{n-2}\omega_n \int_{\tilM} \left[ 
| \nabla_{g_\del} u_\del |^2 - 
 c_n {R_\del}_{-} {u_\del}^2 \right] d g_\del .
\end{equation}
We note that the integral term above approaches $0$ by 
Proposition \ref{supestimate}, (\ref{boundRneg}) and (\ref{IandH}).
Hence, we have that 
$$ \lim_{\del \rightarrow 0} m(\tilg) =
   \lim_{\del \rightarrow 0} m(g_\del) = m(\G).$$ \stop 

Applying the classical PMT \cite{Sch-Yau} to each $\tilg$, we have that 
$m(\tilg) \geq 0$. Thus, the non-negativity of $m(\G)$ follows 
directly from Lemma \ref{massconverge}. 

\subsection{Scalar Curvature Concentration}
In this subsection, we assume that there exists strict jump of mean curvature 
across $\S$, i.e.
$$ H(\S, g_{-})(x) > H(\S, g_{+})(x) \ \ \ \mathrm{for\ some}
   \  x \in \S .$$
We will prove that $\G$ has a strict positive mass.

Since $H(\S, g_{-})$ and $H(\S, g_{+})$ both are continuous functions on $\S$,
we can choose a compact set $K \subset \S$ so that   
\begin{equation}
  H(\S, g_{-})(x) - H(\S, g_{+})(x) \geq \eta, \ \ \forall x \in K
\end{equation} 
for some fixed $\eta > 0$. By Proposition \ref{smoothcorner}, 
we have that
\begin{eqnarray} 
{R_\del}_{+}(x,t) \geq \eta \left\{ \frac{100}{\del^2} 
\phi(\frac{100t}{\del^2}) \right\} - C_0, 
& \forall (x,t) \in K \times [-\frac{\del^2}{100},\frac{\del^2}{100}] ,
\label{Rconcentration}
\end{eqnarray}
which suggests that the scalar curvature of
$g_\del$ and $\tilde{g}_\del$ has a fixed amount of concentration on $K$.

To exploit this fact, we use conformal deformation again to make 
$\tilde{g}_\del$ even scalar flat. Since $\tilR = {u_\del}^\frac{4}{2-n} 
{R_\del}_{+} \geq 0$, $\exists$ a $C^2$ positive solution to the following
equations
\begin{equation}\label{secondconf}
 \left\{ \begin{array}{rrl}
            {\triangle}_{\tilde{g}_\del} v_\del -
              c_n \tilde{R}_\del v_\del  & = & 0 \\ 
            \lim_{x \rightarrow \infty} v_\del & = & 1 .
          \end{array} 
  \right.
\end{equation} 
By the maximum principle, we have that
\begin{equation}\label{boundv}
0 < v_{\del} \leq 1 .
\end{equation}
Now define
\begin{equation}
\hatg = {v_\del}^\frac{4}{n-2} \tilg .
\end{equation}
Similar to previous discussion, we know that $\hat{g}_\del$ is an 
asymptotically flat metric and the scalar curvature of $\hat{g}_\del$ is 
identically zero. Furthermore, $m(\hatg)$ and $m(\tilg)$ are related by
\begin{equation}\label{massdifference2}
 m(\tilg) = m(\hatg) + \frac{n-1}{n-2} \omega_n \int_{\tilM}  \left[ 
| \nabla_{\tilg}  v_\del |^2 + c_n \tilR v_\del^2 \right] \ d\tilg , 
\end{equation}
where $m(\hatg) \geq 0$ by the classical PMT. Hence, to prove $m(\G) > 0$,
it suffices to show the integral term in (\ref{massdifference2}) has a strict 
positive lower bound.

\begin{prop}
\begin{equation}\label{infenergy}
 \inf_{\del > 0} \left\{ \int_{\tilM} \left[ | \nabla_{\tilg} v_\del |^2 + 
 c_n \tilR {v_\del}^2 \right] \ d\tilg \right\} > 0
\end{equation}
\end{prop}  

\pf Assume (\ref{infenergy}) is not true, passing to a subsequence, we may 
assume that
\begin{equation} \label{zeroYenergy}
 \lim_{\del \rightarrow 0} \int_{\tilM} \left[ | \nabla_{\tilde{g}_\del} 
 v_\del |^2 + c_n \tilde{R}_\del {v_\del}^2 \right] \ d\tilg = 0 .
\end{equation}  
Since $\tilR \geq 0$, (\ref{zeroYenergy}) is equivalent to
\begin{equation} \label{zep}
          \lim_{\del \rightarrow 0} 
          \int_{\tilM} | \nabla_{\tilg} v_\del |^2 \ d\tilg = 0 
                    \mathrm{\ \ \ and\ \ } 
          \lim_{\del \rightarrow 0} \int_{\tilM} \tilR {v_\del}^2 
\ d\tilg  =  0 .
\end{equation}
Outside $\S \times [-\frac{\del}{2},\frac{\del}{2}]$, we have $g_\del = g$.
Hence, (\ref{secondconf}) becomes
\begin{equation}\label{vsubharmonic}
\triangle_{\tilde{g}_\del} v_\del - c_n \left(u^\frac{4}{2-n}_\del R(g)_{+}
\right) v_\del = 0 .
\end{equation}
It follows from Proposition \ref{supestimate},  
(\ref{boundv}) and Schauder Estimates that, passing to a 
subsequence, $v_\del$ converges to a function $v$ in $C^2$ topology on compact
sets away from $\S$. By (\ref{zep}), we have that
\begin{equation}
  \int_{\tilM \setminus \S} | \nabla_g v |^2 \ dg = 0 ,
\end{equation}
which shows that $v$ is a constant on $\Omega$ and $\tilM \setminus 
\lbar{\Omega}$.

We claim that $v = 1$ on $\tilM \setminus \lbar{\Omega}$. If not, we may 
assume $v = \beta < 1 $ by (\ref{boundv}). We fix a $\del_0 \in (0, \ep)$ and 
denote the region inside
$\S \times \{\del_0\} $ by $\Omega_{\del_0}$. For each $\del < \del_0$, 
we let 
$w_\del$ be the solution to the following equations
\begin{equation}\label{barrier}
\left\{ \begin{array}{rl}
         \triangle_{\tilde{g}_\del} w_\del = 0 & \mathrm{\ on\  } \tilM 
         \setminus \lbar{\Omega}_{\del_0} \\
         w_\del = v_\del & \mathrm{\ on\  } 
         \S \times \{\del_0\} \\
         w_\del(x) \rightarrow 1  & \mathrm{\ at\  } \infty .
        \end{array} 
\right.
\end{equation}
Since $w_\del$ minimizes the Dirichlet energy among all functions with the 
same boundary values, we have that 
\begin{equation}\label{barrierenergy}
\int_{\tilM \setminus \lbar{\Omega}_{\del_0}} | \nabla_{\tilg} w_\del |^2 
\ d\tilg \leq \int_{\tilM \setminus \lbar{\Omega}_{\del_0}} | 
\nabla_{\tilg} v_\del |^2\  d\tilg .
\end{equation}
On the other hand, if we choose $w$ to solve 
\begin{equation}
\left\{ \begin{array}{rl}
         \triangle_{g} w = 0 & \mathrm{\ on\ } \tilM \setminus
         \lbar{\Omega}_{\del_0} \\
         w = \beta & \mathrm{\ on\ } \S \times \{\del_0\} \\
         w(x) \rightarrow 1  & \mathrm{\ at\ } \infty ,
        \end{array} 
\right.
\end{equation} 
we have that
\begin{equation}\label{limitenergy}
\int_{\tilM \setminus \lbar{\Omega}_{\del_0}} | \nabla_{g} w |^2 \ dg =
\lim_{\del \rightarrow 0} \int_{\tilM \setminus \lbar{\Omega}_{\del_0}} 
| \nabla_{\tilg} w_\del |^2 \ d\tilg ,
\end{equation}
because $\tilg \rightarrow g$ uniformly on $\tilM$ and $ v_\del \rightarrow 
\beta$ uniformly on $\S \times \{ \del_0\}$. Hence, it follows from 
(\ref{zep}), (\ref{barrierenergy}) and (\ref{limitenergy}) that
\begin{equation}
\int_{\tilM \setminus \lbar{\Omega}_{\del_0}} | \nabla_{g} w |^2 dg \leq 
\lim_{\del \rightarrow 0} \int_{\tilM \setminus \lbar{\Omega}_{\del_0}} 
| \nabla_{\tilde{g}_\del} v_\del |^2 d\tilde{g}_\del = 0  ,
\end{equation}
which implies that $w$ must be a constant. Since $\beta < 1$, we get a 
contradiction. Therefore, $v = 1$ on $\tilM \setminus \lbar{\Omega}$.

Next, we let $\mu$, $\mu_\del$ denote the $(n-1)$-dimensional volume measure
induced by $g$, $\tilg$ on $\S$ and let $e_\del$ denote the energy 
$\int_{\tilM} |\nabla_{\tilg} v_\del|^2 \ d\tilg$.

We fix $0 < \theta < 1$ and $0 < \sigma < \ep$. Since $v_\del \rightarrow
1$ uniformly on compact set away from $\S$, we have that
\begin{equation}\label{vlessone}
  v_\del > \theta \mathrm{\ on \  } \S_\sigma, \ \mathrm{\ for\ } \del 
\ll 1 ,
\end{equation}
where $\S_t$ is the slice $\S \times \{t\}$.
We do all the estimates inside the strip $N_\sigma = \S \times
[-\sigma, \sigma]$. First, we have that
\begin{equation}\label{lestimate1}
\int_{\S} \left\{ \int^{\sigma}_{-\sigma} | \nabla_{\tilg} v_\del(x,t)
 |^2 dt \right\} \ d{\mu_\del(x)} \leq C_1 \int_{N_\sigma} | \nabla_{
 \tilg} v_\del |^2 \ d\tilg \leq C_1 e_\del .
\end{equation}
Let $l_\del(x) =  \int^{\sigma}_{-\sigma} | \nabla_{\tilg} 
v_\del(x,t) |^2 dt $, (\ref{lestimate1})  becomes
\begin{equation}\label{lestimate2}
\int_{\S} l_\del(x) \ d{\mu_\del(x)} \leq  C_1 e_\del .
\end{equation}
For any $k > 1, \del >0 $, we define
\begin{eqnarray}
  A_{\del, k}  =  \left\{ x \in \S \left|  l_\del(x) \leq k
  \frac{C_1 e_\del}{\mu_\del(\S)} \right. \right\} \\
  A^K_{\del, k}  =  A_{\del, k} \cap K \\
  A^K_{\del, k, \sigma}  =  A^K_{\del, k} \times [-\sigma, \sigma] .
\end{eqnarray}  
By (\ref{lestimate2}) we have that
\begin{equation}\label{measuredel}
  \mu_\del (A_{\del,k} ) \geq ( 1 - \frac{1}{k}) \mu_\del(\S) .
\end{equation} 
Since $\mudel$ is uniformly close to $\mu$, (\ref{measuredel}) implies 
that
\begin{equation}
 \mudel(A^K_{\del, k}) \geq \frac{1}{2} \mudel(K)
\end{equation}
for some fixed large $k$ and any $\del \ll 1$.
 
For any $(x, t) \in A^K_{\del, k, \sigma}$, we have that 
\begin{eqnarray}
 | v_\del(x,\sigma) - v_\del(x,t) | & \leq & C_2 \int^{\sigma}_{-\sigma} 
 | \nabla_{\tilde{g}_\del} v_\del | (x,t) dt \nonumber \\
  & \leq & C_2 (2\sigma)^\frac{1}{2} \left\{\int^{\sigma}_{-\sigma} 
 | \nabla_{\tilg} v_\del |^2 (x,t) dt \right\}^\frac{1}{2} \nonumber \\
  & = & C_2 (2\sigma)^\frac{1}{2} l_\del(x)^\frac{1}{2} \nonumber \\
  & \leq & C_2 (2\sigma)^\frac{1}{2} \left\{k \frac{C_1 e_\del}{\mudel(\S)} 
  \right\}^\frac{1}{2} .
\end{eqnarray}
It follows from (\ref{vlessone}) that
\begin{equation}
 v_\del(x,t) \geq \theta - C_2 (2\sigma)^\frac{1}{2} \left\{ k 
 \frac{C_1 e_\del}{\mudel(\S)} \right\}^\frac{1}{2} .
\end{equation} 
On the other hand, for $x \in A^K_{\del,k}$, we have that
\begin{equation}
\int^{\del}_{-\del} \tilR(x,t) dt \geq u^{\frac{4}{2-n}}_\del(x) 
\int^{\frac{\del^2}{100}}_{-\frac{\del^2}{100}} \left\{ 
 \eta \left\{ 
  \frac{100}{\del^2}\phi(\frac{100t}{\del^2}) \right\} - C_0 \right\} dt .
\end{equation}
Therefore, we have the following estimate
\begin{eqnarray}
 \liminf_{\del \rightarrow 0} \int_{A^K_{\del,k, \sigma}} \tilR {v_\del}^2 \ 
 d\tilg \geq  \nonumber \\ \liminf_{\del \rightarrow 0} \left\{ \left\{ \theta
  - C_2 \left\{ (2\sigma) k \frac{C_1 e_\del}{\mudel(\S)} \right\}^\frac{1}{2} 
 \right\}^2 \int_{A^K_{\del,k, \sigma}} \tilR \ d\tilg \right\} \geq \nonumber
 \\
 \theta^2 C_3 \liminf_{\del \rightarrow 0} \left\{ \int_{A^K_{\del,k}} \left\{
 \int^{\del}_{-\del} \tilR(x,t) dt \right\} d\mudel \right\} \geq \nonumber \\
 C_3 \theta^2 \eta \liminf_{\del \rightarrow 0} \mudel(A^K_{\del,k})  \geq 
 \nonumber \\
 \frac{1}{2} C_3 \theta^2 \eta \mu(K) > 0 
\end{eqnarray}
which is a contradiction to (\ref{zep}). \stop

We conclude that $\G$ has a strict positive mass in case there exists strict 
jump of mean curvature across $\S$. 

\section{Zero Mass Case}

Let $\G=(g_{-}, g_{+})$ satisfy all the assumptions in 
Theorem \ref{PMT1} and $m(\go) = 0$.
The following corollary on $R(\go), R(\gi)$ follows directly from 
Theorem \ref{PMT1}.

\begin{corol} \label{0rg}
Under the above assumptions, $g_{-}$ and $g_{+}$ both have zero 
scalar curvature in $\Omega$ and $M \setminus \lbar{\Omega}$. 
\end{corol}

\pf First, we assume that $R(g_{-})$ is not identically zero in $\Omega$. Let 
$u$ be a positive solution to the equation
\begin{equation}
 \left\{ \begin{array}{rl}
         \triangle_{g_{-}} u - c_n R(g_{-}) u = 0 &  \mathrm{\ \ on\ } 
         \Omega \\
         u = 1 &  \mathrm{\ \ on\ } \S  .     
         \end{array}
 \right.
\end{equation}
Consider $\tilde{\G}= (\tilde{g}_{-}, g_{+})$, where $\tilde{g}_{-} = u^
{\frac{4}{n-2}} g_{-}$. Since $u$ solves the conformal Laplacian of $\gi$, 
$\tilde{g}_{-}$ has zero scalar curvature. By the strong maximum 
principle, we have $\frac{\partial u}{\partial \nu} > 0$, where $\nu$ is the 
unit outward normal to $\S$. A direct computation shows that
\begin{equation}
H(\S, \tilde{g}_{-})
(x) = H(\S, g_{-})(x) + \frac{2}{n-2} \frac{\partial u}{\partial \nu}(x) .
\end{equation} 
Hence, $H(\S, \tilde{g}_{-}) > H(\S, g_{-}) \geq H(\S, g_{+})$. Applying 
Theorem \ref{PMT1} to $\tilde{\G}$, we see that $m(\tilde{\G}) > 0$,
which is a contradiction.

Second, we assume that $R(g_{+})$ is not identically zero in $M \setminus 
\lbar{\Omega}$. Let $v$ be a positive solution to 
\begin{equation}
 \left\{ \begin{array}{rl}
           \triangle_{g_{+}} v - c_n R(g_{+}) v = 0 & \mathrm{\ \ on\ }
            M \setminus \lbar{\Omega} \\
           v = 1 & \mathrm{\ \ on\ } \S \\
           v \rightarrow 1 &  \mathrm{\ \ at\ } \infty .
         \end{array} 
 \right.
\end{equation}
Consider $\hat{\G}= (g_{-}, \hat{g}_{+})$, where $\hat{g}_{+} = v^
{\frac{4}{n-2}} g_{+}$. A similar argument shows that $\hat{g}_{+}$ is 
scalar flat in $M \setminus \lbar{\Omega}$ and 
$H(\S, \hat{g}_{+}) < H(\S, g_{+}) \leq H(\S, g_{-}). $
Therefore, Theorem \ref{PMT1} implies that $m(\hat{\G}) > 0$. On the other 
hand, we have that 
\begin{equation}
m(\hat{\G}) = m(\G) + A ,
\end{equation}
where $ v = 1 + A |x|^{2-n} + O(|x|^{1-n})$. 
By the maximum principle, $ A \leq 0$. Hence, $m(\G) \geq m(\hat{\G})> 0$, 
which is again a contradiction to the assumption that $m(\G) =0$. \stop

Corollary \ref{0rg} only reveals information on the scalar curvature, it would
be more interesting to know if $m(\G)=0$ implies that $\G$ is flat away from 
$\S$. Such a type of questions has been studied by 
H. Bray and F. Finster in \cite{Bray-Finster}.
In particular, they obtained the following result concerning the mass
and the curvature of a metric which can be approximated by smooth 
metrics in their sense. 

\begin{prop}{\rm \cite{Bray-Finster}} \label{bray-finster}
Suppose $\{ g_i \}$ is a sequence of $C^3$, complete, asymptotically flat 
metrics on $M^3$ with non-negative scalar curvature and the total masses 
$\{ m_i \}$ which converge to a possibly non-smooth limit metric $g$ in the 
$C^0$ sense. Let $U$ be the interior of the sets of points where this 
convergence of metrics is locally $C^3$.

Then if the metrics $\{ g_i \}$ have uniformly positive isoperimetric
constants and their masses $\{ m_i \}$ converges to zero, then $g$ is flat in 
$U$.
\end{prop}  

Now we are in a position to show that, in case 
$n = 3$, $\G$ is regular cross $\S$ and $(M, \G)$ is isometric 
to $(\R^3,g_o)$.

\vspace{.2cm}

\noindent \emph{Proof of Theorem \ref{PMT2}}: 
First, we show that $\gi$ and $\go$ are flat in $\Omega$ and 
$M \setminus \lbar{\Omega}$.
Since $\gi$ and $\go$ are $C^{3, \alpha}_{loc}$,
it follows from the proof of Proposition \ref{supestimate} that $\{ \tilg \}$ 
converges to $g$ locally in $C^3$ away from $\S$. By Proposition 
\ref{smoothcorner}, we know that $\tilg$ and $g$ are uniformly 
close on $\tilM$, hence $\{ \tilg \}$ has uniformly 
positive isoperimetric constants. By Lemma \ref{massconverge}, we know that 
$\lim_{\del \rightarrow 0} m(\tilg) = 0$. Therefore, $g_{-}$ and  $g_{+}$
are flat by Proposition \ref{bray-finster}. 

Second, we show that $A_{-} = A_{+}$, where $A_{-}$ and 
$A_{+}$ are the second fundamental forms of $\S$ in $(\lbar{\Omega}, \gi)$
 and $(M \setminus \Omega, \go)$. Taking trace of the Codazzi equation 
and using the fact that $\gi, \go$ is flat, we have that
\begin{equation}
\left\{
\begin{array}{ccl}
 div_{g_\sigma} A_{-} & =  & \nabla H(\S, \gi) \\
 div_{g_\sigma} A_{+} & =  & \nabla H(\S, \go) ,
\end{array}
\right.
\end{equation}
where $g_\sig$ is the induced metric $\gi |_{\S} = \go |_{\S}$.
On the other hand, Theorem \ref{PMT1} implies that 
$H(\S, \gi) \equiv H(\S, \go)$ on $\S$. Hence, 
\begin{equation}
  div_{g_\sigma} (A_{-} - A_{+}) = 0 \ \ 
  \mathrm{and\ \ }  
  tr_{g_\sigma} (A_{-} - A_{+}) = 0.
\end{equation}
We recall the fact that any divergence free and trace free
$(0,2)$ symmetric tensor on $(S^2, g_{\sigma})$ 
must vanish identically \cite{Hopf}, thus we conclude that $A_{-} = A_{+}$.
Now it follows from the fundamental theorem of surface theory in $\R^3$ that 
$\G$ is actually $C^2$ across $\S$. The classical 
PMT \cite{Sch-Yau} then implies that $(M, \G)$ is isometric to $\R^3$ with 
the standard metric. \stop

\bigskip
\noindent{\bf Acknowledgments:}
I would like to thank my Ph.D. advisor Professor Richard Schoen for bringing 
up this problem and for his superb direction. I also would like to thank 
Professor Robert Bartnik and Professor Hubert Bray for many stimulating 
discussions.

\vspace{.2cm}

\bibliographystyle{plain}
\bibliography{PMT}

\end{document}